\documentclass[epj]{webofc}
\usepackage[utf8]{inputenc}
\usepackage[varg]{txfonts}   

\usepackage{booktabs}
\usepackage{xcolor}
\definecolor{darkred}{rgb}{0.4,0.0,0.0}
\definecolor{darkgreen}{rgb}{0.0,0.4,0.0}
\definecolor{darkblue}{rgb}{0.0,0.0,0.4}
\usepackage[bookmarks,linktocpage,colorlinks,
    linkcolor = darkred,
    urlcolor  = darkblue,
    citecolor = darkgreen]{hyperref}
\usepackage{subfigure}
\usepackage{amsmath}
\wocname{EPJ Web of Conferences}
\woctitle{Lattice2017}
\begin{document}
%
\selectlanguage{english}
%
%
%
\title{%
Charmonium resonances on the lattice
}
\author{%
\firstname{Gunnar} \lastname{Bali}\inst{1} \and
\firstname{Sara} \lastname{Collins}\inst{1} \and
\firstname{Daniel} \lastname{Mohler}\inst{4,5} \and
\firstname{M.} \lastname{Padmanath}\inst{1} \and 
\firstname{Stefano} \lastname{Piemonte}\inst{1}\fnsep\thanks{Speaker, \email{stefano.piemonte@physik.uni-regensburg.de}} \and
\firstname{Sasa} \lastname{Prelovsek}\inst{1,2,3} \and
\firstname{Simon} \lastname{Weishäupl}\inst{1}\fnsep\thanks{Speaker, \email{simon.weishaeupl@physik.uni-regensburg.de}}
}
\institute{%
Institute for Theoretical Physics, University of Regensburg, 93040 Regensburg, Germany
\and
Faculty of Mathematics and Physics, University of Ljubljana, 1000 Ljubljana, Slovenia
\and
Jozef Stefan Institute, 1000 Ljubljana, Slovenia
\and
Helmholtz-Institut Mainz, 55099 Mainz, Germany
\and
Johannes Gutenberg Universit\"at Mainz, 55099 Mainz, Germany
}
\abstract{The nature of resonances and excited states near decay
  thresholds is encoded in scattering amplitudes, which can be extracted from
  single-particle and multiparticle correlators in finite
  volumes. Lattice calculations have only recently reached the
  precision required for a reliable study of such correlators. The
  distillation method  represents a significant improvement insofar as
  it simplifies quark contractions and allows one to easily extend the
  operator basis used to construct interpolators. We present
  preliminary results on charmonium bound states and resonances on the
  $N_f=2+1$ CLS ensembles. The long term goal of our investigation is to
  understand the properties of the X resonances that do not fit into
  conventional models of quark-antiquark mesons. We tune various
  parameters of the distillation method and the charm quark mass. As a first
  result, we present the masses of the ground and excited states in the $0^{++}$ and $1^{--}$ channels.  }
\maketitle
\section{Introduction}\label{intro}

The mysterious properties of exotic QCD resonances have motivated an
increasing interest in non-perturbative lattice calculations during
the last decade. Monte-Carlo simulations of QCD regularized on the
lattice can provide an alternative insight into the fundamental
mechanisms of strong interactions, but the study of resonances is a
challenge that requires a huge numerical and theoretical effort. In
particular, scattering of particles cannot be investigated directly on
an Euclidean space-time lattice, but, following the L\"uscher method \cite{Luscher},
scattering amplitudes are instead extracted from the finite volume dependence
of the discrete energy levels of the spectrum of two interacting
particles in a box. Many simulations with either different volumes or in
different moving frames are needed and a large number of configurations is required to measure the
lattice energy levels with a sufficient precision to compute scattering
observables. Only recently lattice QCD calculations have been able to reach
sufficiently small lattice spacings and pion masses to allow for a reliable
study of QCD resonances with heavy flavor content.

The aim of our project is the understanding of the nature of
charmonium(-like) resonances and excited states near decay
thresholds by means of non-perturbative lattice simulations. In this
contribution we present our recent progress, regarding the calculation
of the energy levels on the $N_f=2+1$ ensembles
generated by the CLS collaboration. We employ the distillation method,
which significantly improves the computation of all diagrams involving
single-meson and two-meson correlation functions. In Section
\ref{explat} we present the current experimental status, the results
of previous lattice investigations and the motivations for our
project. In Section \ref{lattice_setup} we describe the details of our
ensembles and the tuning of the charm quark mass, while in Section
\ref{distillation} we give an overview of the distillation method
and the parameters used in our analysis. Our results are presented
in Section \ref{results} and are followed by the conclusions in Section \ref{conc}.

\section{Experimental status and previous lattice investigations}\label{explat}

Our final goal is to consider conventional and exotic charmonium(-like) resonances. 
Here we focus on the vector and scalar channels.  The   
resonances are labeled ``XY'' states, the
letter $Y(\dots)$ is used to denote the states with quantum numbers
$1^{--}$, while the letter $X(\dots)$ is used for all other flavorless
charmonium exotic resonances.

The two lightest states of the charmonium spectrum in the vector
channel, the $J/\psi$ and $\psi(2S)$, are relatively stable and decay
by charm-quark annihilation or weak interactions. The third state, the
$\psi(3770)$, is a resonance which decays by strong interactions, lying above the
$\bar{D}D$ decay threshold. The decay modes of the $\psi(3770)$ are
in agreement with those of an excitation of a quark-antiquark meson. At higher energies,
there are many other states whose structure and constituents are
unknown, like the $Y(4260)$, where BESIII recently found two peaks $Y(4240)$ and 
$Y(4320)$ \cite{Ablikim:2016qzw}. The higher peak is consistent with   $Y(4360)$ reported earlier. 
Our first step toward the study of these exotic states is to
investigate the resonance $\psi(3770)$, whose properties are
well-known experimentally and that can be used as a benchmark to
estimate carefully the systematic uncertainties of our computation of the
phase shift.

The $\chi_{c0}$ is the lightest bound state in the charmonium spectrum
with quantum numbers $0^{++}$. The higher lying $X(3915)$ was
discovered by Belle in 2004 in $J/\psi \, \omega$ decays \cite{X3915}
and was subsequently listed as the first excited state $\chi_{c0}(2P)$ by
the Particle Data Group \cite{PDGOld}. This identification was based on a
determination of its quantum numbers by BaBar \cite{Lees:2012xs}, and
has since been challenged. The OZI rule favors an excited $\bar{c}c$
state decaying into $D\bar{D}$ mesons, but not via $J/\psi \, \omega$
\cite{X3915r1} and, at present, there is still no evidence for a
resonance at 3915 MeV in the s-wave $D\bar{D}$ meson scattering
channel. Considering ``exotic'' interpretations of the state, a
$\bar{c}c$-gluon exotic hybrid scenario has been excluded by lattice
calculations as such states appear to be significantly heavier than
all other low energy resonances \cite{Liu}. The suppression of
$D\bar{D}$ meson decays could instead be explained easily if the
$X(3915)$ has a hidden strange content. A possible structure could be
in this case a bound state of two $D_s$ mesons, and the 20~MeV gap of
$X(3915)$ from the $\bar{D}_s D_s$ decay threshold would be
interpreted as the binding energy of the meson molecule. The decays
into $J/\psi \, \omega$ could be explained in terms of the
$\omega-\phi$ mixing \cite{X3915r4}. Even in this case it is still
difficult to explain why the $X(3915)$ has not yet been seen in the
$\eta \eta_c$ decay channel, given the larger mixing in pseudoscalar
light-strange sector with respect to the vector case
\cite{X3915r3}. An appealing hypothesis is that the $X(3915)$ is a
$cs\bar{c}\bar{s}$ tetraquark, i.e. a bound state with four valence
quarks without a definite internal structure. 

Another solution of this puzzle has been suggested in \cite{Zhou:2015uva}, where the
authors argue that BaBar made an unjustified assumption in their analysis,
which results in the exclusion of $J^{PC}=2^{++}$ as the quantum numbers for
the $X(3915)$. Indeed the authors argue that the $X(3915)$ is simply the
already known $\chi_{c2}(2P)$ \cite{PDG}. If the
$X(3915)$ is not the first excited state of the $\chi_{c0}$, there is
the question of what the energy of such a state could be. The
Belle collaboration has recently reported a new scalar charmoniumlike
state $X^{\ast}(3860)$ that decays into $D\bar{D}$ mesons as expected
for the excited state of the $\chi_{c0}$ \cite{Chilikin:2017evr}.

The lack of a deep knowledge of the scalar and vector exotic
charmonium resonances is a strong motivation for non-perturbative
lattice calculations of the scattering matrix. Previous investigations
performed by some of us have been able to extract the elastic scattering phase shift
from a basis including both quark-antiquark and $D\bar{D}$ interpolators in the scalar
channel~\cite{Lang:2015sba}. However, the study of the resonance
structure did not resolve the issue of the first excited scalar charmonium. These exploratory results, which 
did not consider strange quarks in the valence or the sea sector,  gave an indication 
for a bound state $\chi_{c0}$ and a rather narrow resonance slightly below $4~$GeV.  More simulations are
required to reach a conclusive picture of the scalar resonances,  particularly
to address the experimental  $X^{\ast}(3860)$ and $X(3915)$ states. We plan to extend the previous analysis by considering also a coupling to a hidden strange sector. This will entail coupled  scattering of $\bar DD$ and $\bar D_sD_s$ channels.  

\section{Lattice Setup}\label{lattice_setup}

\begin{figure}[tp]
	\centering
	\includegraphics[scale=0.43]{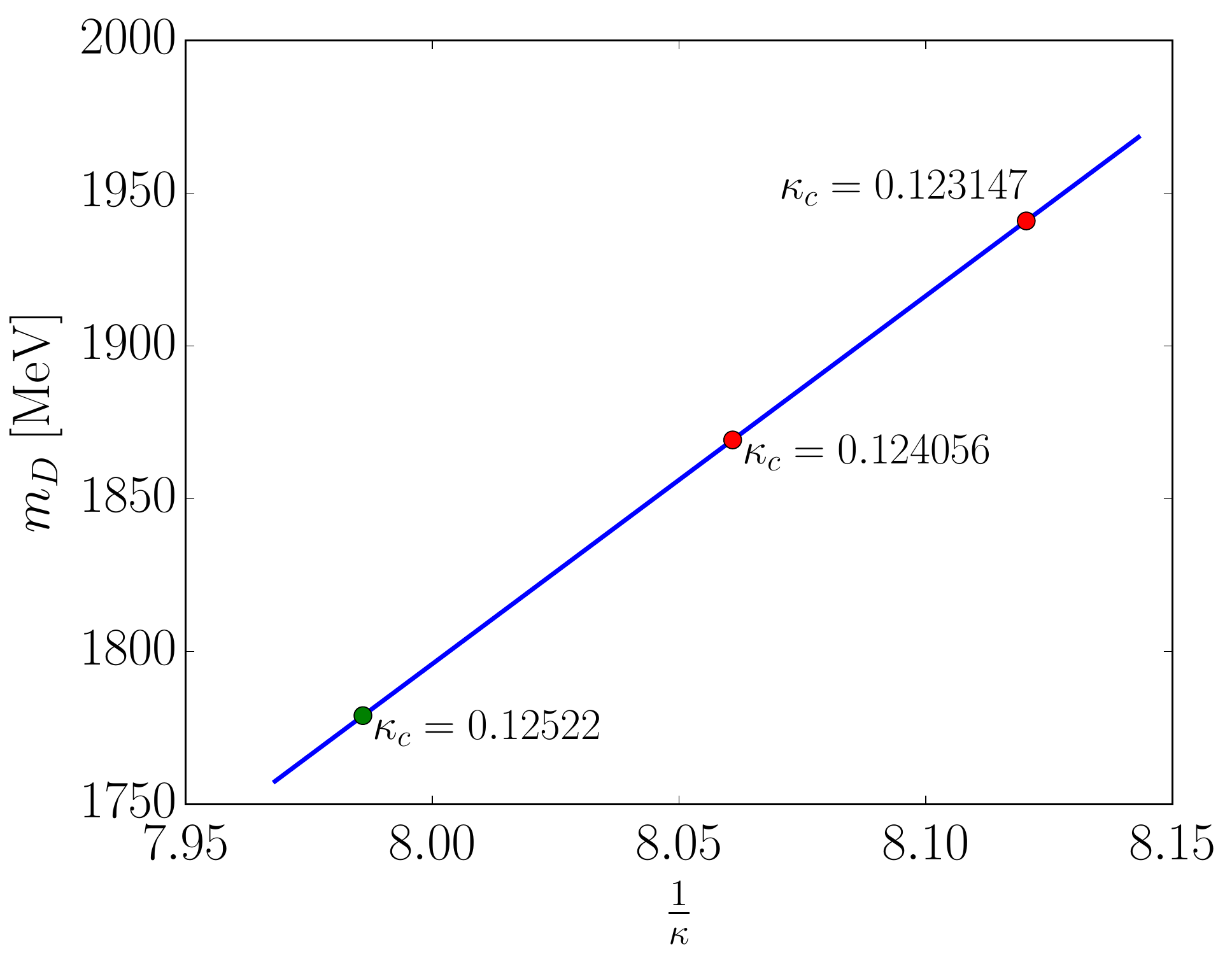}
	\caption{Dependency of the D-meson mass~$m_D$ on the bare
          charm quark mass~$m$ for different \mbox{$\kappa_c$ -
            values.} The two red points indicate the previously tuned
          values of $\kappa_c$. The blue line indicates a linear fit
          to determine the new value \mbox{$\kappa_c = 0.125220$}
          corresponding to a D-meson with a mass 80 MeV lighter than its
          physical value.}
	\label{fig:kappa_fit}
\end{figure}

We perform our analysis on the $N_f=2+1$ ensembles generated by the
CLS collaboration, see, for example, Refs.~\cite{Bruno:2014jqa} and
\cite{Bali:2016umi} for details. Two ensembles have been so far
considered, labeled as U101 and H105, both with $m_\pi=280$ MeV and
lattice spacing $a=0.0854$ fm, but with different lattice volumes of
$24^3\times 128$ and $32^3\times 96$, respectively. Our lattice
discretization of the QCD action includes non-perturbatively
$\mathcal{O}(a)$-improved Wilson fermions with tree-level Symanzik
improved gauge action. Open boundary conditions have been applied to
the fermion and gluon fields in the time direction, while they are periodic in space. 

The strange and the light quarks are dynamical and their masses are
fixed in the Monte-Carlo simulation, however, we have the freedom to
choose the mass of the (quenched) charm quark. The hopping parameter
$k_c$, where the bare quark mass $am_c=(1-8k_c)/(2k_c)$, is usually
fixed so that the experimental value of the spin average of the
$J/\psi$ and the $\eta_c$ meson masses is reproduced. For the current
project, for which we employ ensembles with unphysical pion masses,
this may result in the $\psi(3770)$ lying below the $D\bar{D}$
threshold, so that the s-wave decay into $D\bar{D}$ mesons would be
forbidden.  This scenario might be avoided by reducing the charm quark
mass. We therefore use two different $k_c$ interpolating between a $D$
meson mass approximatively 80 MeV above and below its physical value
to try estimate how the energy levels move with respect to the
threshold as a function of the charm quark mass, see
Fig.~\ref{fig:kappa_fit}.

\section{Distillation method and variational analysis}\label{distillation}

To study the charmonium spectrum, we use the distillation method
(quark-field smearing) \cite{Peardon:2009gh} and the variational
analysis of the correlation matrix between single and multi-particle
operators at different timeslices. The wave function is constructed
from the eigenvectors of the 3D Laplace operator $\nabla^2_t$ defined
on a given timeslice $t$. A fermion wave function defined on a single
lattice point can be smeared by the ``distillation'' operator
\begin{equation}
\Box (t) = \sum_{i=0}^N \Psi_i(t) \Psi_i^\dag(t) \,,
\end{equation}
where the sum runs over the eigenvectors $\Psi_i (t)$ of $\nabla^2_t$
corresponding to the $N$ lowest eigenvalues. The
resulting wave function has a Gaußian bell shape, whose width is
controlled by the number of eigenvectors included in the sum,
typically of the order of $O(100)$. If $N$ is taken as large as $3\times
V_3$, to span the entire linear space, then the distillation operator
reduces to the identity operator and no smearing is applied at all.

\begin{figure}
 \subfigure[Connected and disconnected pseudoscalar correlators]{\includegraphics[width=0.49\linewidth]{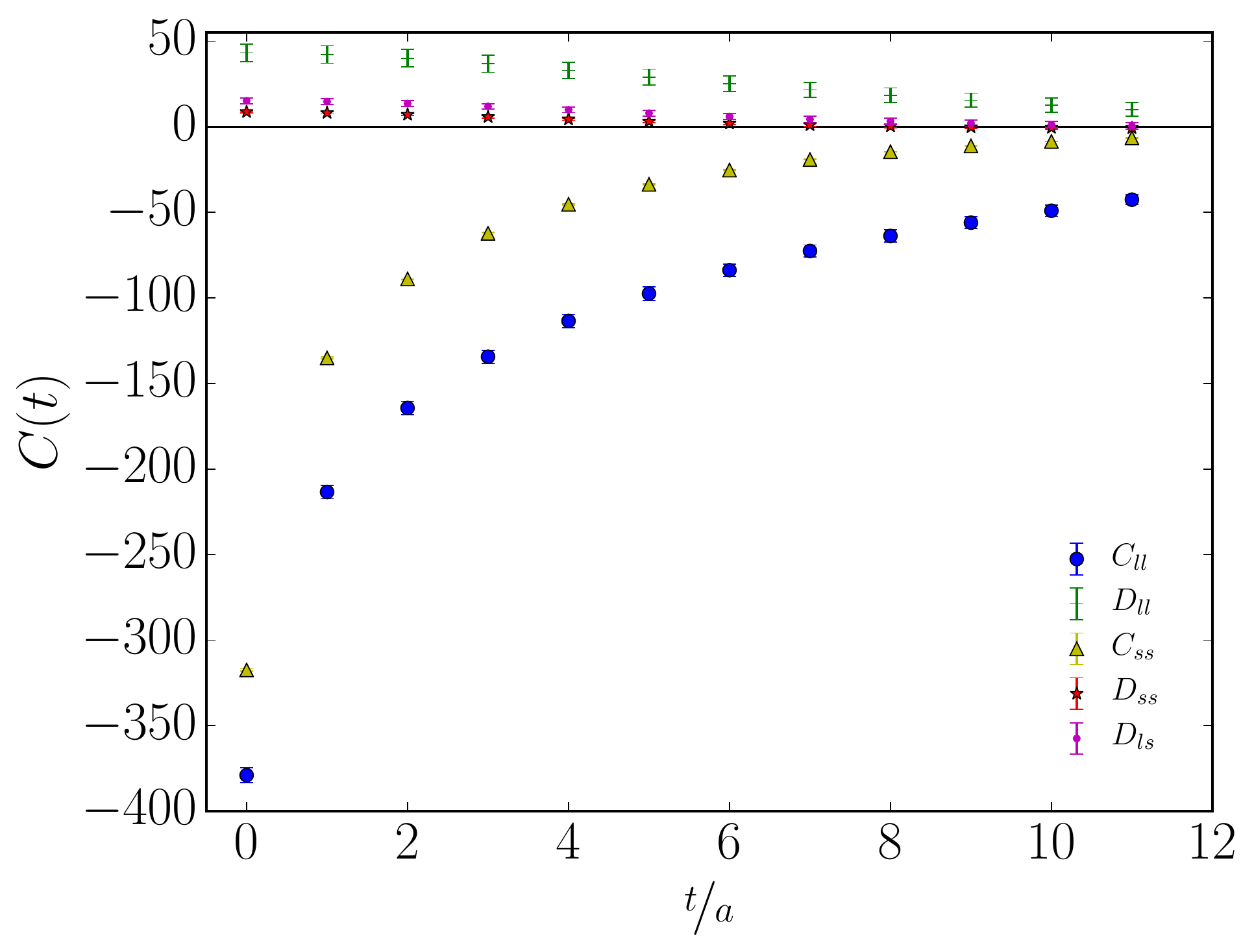}\label{fix_nev_b}}
 \subfigure[Effective mass of the first excited state of the $\eta_c$]{\includegraphics[width=0.49\linewidth]{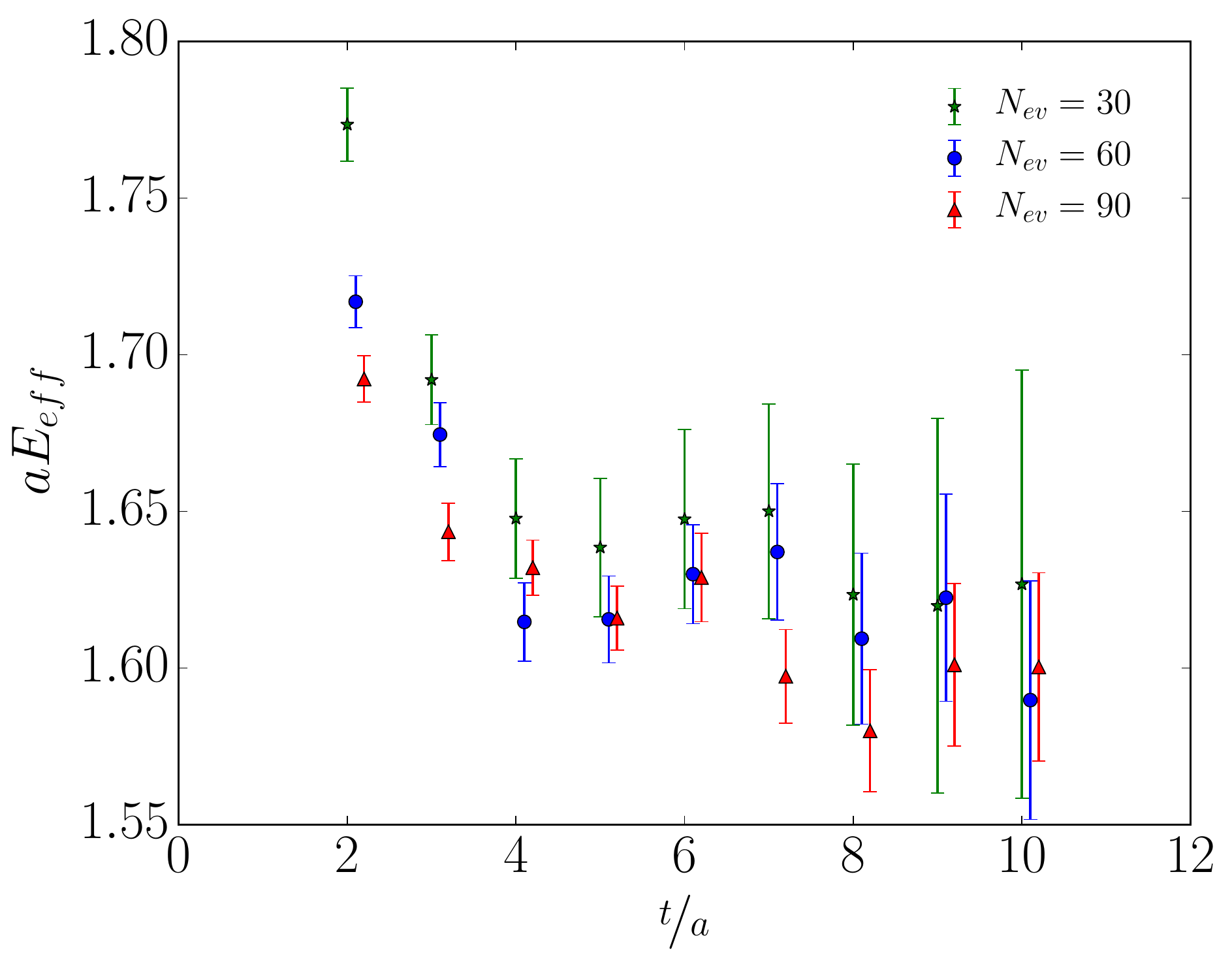}\label{fix_nev_a}}
 \caption{ \ref{fix_nev_b} Trace of the contribution to the pseudoscalar correlator for
   strange and light quarks (related to $\eta$ and $\eta^\prime$ mesons). The disconnected contributions
   $D_{ff}$ for the dynamical flavors $f$ are smaller in magnitude but
   have a larger statistical uncertainty than the connected
   contributions $C_{ff}$. \ref{fix_nev_a} Effective mass plots for the first excited charmonium
   state ($0^{-+}$) using 30, 60 and 90 Laplacian eigenvectors.}
\end{figure} 

In the distillation method, a generic two-point correlation function
can be rewritten as a product of a combination of at least four
matrices
\begin{equation}\label{corr}
C(t,t_0) = \textrm{Tr} \left[ \phi^B(t) \tau(t,t_0) \phi^A(t_0) \tau(t_0,t) \right] \,.
\end{equation}
The perambulators $\tau$, describing the fermion propagation between eigenvectors at possibly different timeslices
\begin{equation}
\tau(t,t_0,i,j)_{\alpha \beta} = \Psi_i^\dag(t) D^{-1}_{\alpha \beta} \Psi_j(t_0)\,,
\end{equation}
are precomputed for the charm, the light and the strange quarks. The total number of inversions required is $4\times N$. Afterwards, the perambulators are combined with the $\phi's$ 
\begin{equation}
\phi^A(t,i,j)_{\alpha \beta} = \Psi_i^\dag(t)  \Gamma^A(t)_{\alpha \beta}\Psi_j(t)\,,
\end{equation}
that carry all information about source/sink operators and quantum
numbers. Finally, the matrices $\phi$ and $\tau$ are multiplied and
traced accordingly to Eq.~\ref{corr} to construct single-meson
correlators and the more involved expressions are required for multi-meson
correlators. We always include diagrams with disconnected or
backtracking strange and light loops, neglecting such contributions
only for the charm. The disconnected contributions for the light
quarks are a typical source of large statistical fluctuations in the
correlation functions, see Fig.~\ref{fix_nev_b}.

The first step of the distillation method is to choose the number of
eigenvectors $N$. The optimal width of the wave function for charmonium
states is expected to be smaller than that of light mesons. One therefore needs
a larger number of eigenvectors compared to
similar analyses of light resonances. Our strategy for searching for the
optimal value of $N$ is to look for the point where an increase of $N$
does not result to a reduction in the error of the
first excited state of the $\eta_c$, see Fig.~\ref{fix_nev_a}. The
number of eigenvectors $N$ has been set to 90 for the U101 ensemble
and to 150 for the H105.

\begin{table}[t]
\caption{Interpolating operators for the $1^{--}$ (left) and the
  $0^{++}$ (right) channel. All repeated indexes are summed over the
  spatial components and derivatives are implemented in a symmetric
  form.}\label{operators}
\begin{minipage}[!t]	{0.5\linewidth}
\begin{tabular}{cc}
	\multicolumn{2}{c}{$T^{--}_1\ $ ($J^{PC} = 1^{--}, 3^{--}, 4^{--}, \dots$)}	\\
	\multicolumn{2}{c}{ }	\\
	Label & Operator \\
	\hline
	0 & $ \bar{q} \ \gamma_i \ q $ \\
	1 & $ \bar{q} \ \gamma_i \gamma_t \ q $ \\
	2 & $ \bar{q} \  \overrightarrow{\nabla}_i \ q $ \\
	3 & $ \bar{q} \ \epsilon_{ijk} \gamma_j \gamma_5 \overrightarrow{\nabla}_k \ q $ \\
	4 & $ \bar{q} \ \overleftarrow{\nabla}_k \gamma_i \overrightarrow{\nabla}_k \ q  $ \\
	5 & $ \bar{q} \ \overleftarrow{\nabla}_k \gamma_i \gamma_t \overrightarrow{\nabla}_k \ q $ \\
	6 & $ \bar{q} \ \overleftarrow{\Delta} \gamma_i \overrightarrow{\Delta} \ q  $ \\
	7 & $ \bar{q} \ \overleftarrow{\Delta} \gamma_i \gamma_t \overrightarrow{\Delta} \ q $ \\
	8 & $ \bar{q} \ \overleftarrow{\Delta} \overrightarrow{\nabla}_i \ q $\\
	9 & $ \bar{q} \ \overleftarrow{\Delta} \epsilon_{ijk} \gamma_j \gamma_5 \overrightarrow{\nabla}_k \ q $\\
	10 & $ \bar{q} \ |\epsilon_{ijk}|  \gamma_j \overrightarrow{D}_k \ q $ \\
	11 & $ \bar{q} \ |\epsilon_{ijk}|  \gamma_j \gamma_t \overrightarrow{D}_k \ q $ \\
	12 & $O^{\bar{D}(-1)D(1)}  \sim \bar{c}\gamma_5l ~\bar{l}\gamma_5 c$ \\
	13 & $O^{\bar{D}(-1)D(1)} \sim \bar{c}\gamma_5\gamma_t l ~\bar{l}\gamma_5 \gamma_t c$\\
	& \\
	&\\
\end{tabular}
\end{minipage}
\hspace{1cm}
\begin{minipage}[!t]{0.5\linewidth}
\begin{tabular}{cc}
	\multicolumn{2}{c}{$A^{++}_1\ $ ($J^{PC} = 0^{++}, 4^{++}, \dots$)}	\\
	\multicolumn{2}{c}{ }	\\
	Label n & Operator \\
	\hline
	0 & $ \bar{q} \ q $ \\
	1 & $ \bar{q} \ \gamma_i \overrightarrow{\nabla}_i \ q $ \\
	2 & $ \bar{q} \ \gamma_i \gamma_t \overrightarrow{\nabla}_i \ q $ \\
	3 & $ \bar{q} \ \overleftarrow{\nabla}_i \overrightarrow{\nabla}_i \ q $ \\
	4 & $ \bar{q} \ \overleftarrow{\Delta} \overrightarrow{\Delta} \ q $ \\
	5 & $ \bar{q} \ \overleftarrow{\Delta} \gamma_i \overrightarrow{\nabla}_i \ q $ \\
	6 & $ \bar{q} \ \overleftarrow{\Delta} \gamma_i \gamma_t \overrightarrow{\nabla}_i \ q $ \\
	7 & $O^{\bar{D}(0)D(0)}  \sim \bar{c}\gamma_5l ~\bar{l}\gamma_5 c$ \\
	8 & $O^{\bar{D}(0)D(0)} \sim \bar{c}\gamma_5\gamma_t l ~\bar{l}\gamma_5 \gamma_t c$\\
	9 & $O^{\bar{D}(p)D(-p)}  \sim \bar{c}\gamma_5l ~\bar{l}\gamma_5 c$\\
	10 & $O^{\bar{D}^{\ast}(0)D^{\ast}(0)}  \sim \bar{c}\gamma_i l ~\bar{l}\gamma_i c$ \\
	11 & $O^{\bar{D}^{\ast}(0)D^{\ast}(0)}  \sim \bar{c}\gamma_i \gamma_t l ~\bar{l}\gamma_i \gamma_t c$ \\
	12 & $O^{J/\psi(0)\omega(0)}  \sim \bar{c}\gamma_i c ~\bar{l}\gamma_i l$ \\
	13 & $O^{J/\psi(0)\omega(0)}  \sim \bar{c}\gamma_i \gamma_t c ~\bar{l}\gamma_i \gamma_t l$ \\
	14 & $O^{\bar{D}_s(0)D_s(0)}  \sim \bar{c}\gamma_5 s ~\bar{s}\gamma_5 c$ \\
	15 & $O^{\bar{D}_s(0)D_s(0)} \sim \bar{c}\gamma_5\gamma_t s ~\bar{s}\gamma_5 \gamma_t c$\\
\end{tabular}
\end{minipage}%
\end{table}

\begin{figure}[t]
\subfigure[Vector channel]{\includegraphics[width=0.55\textwidth,clip]{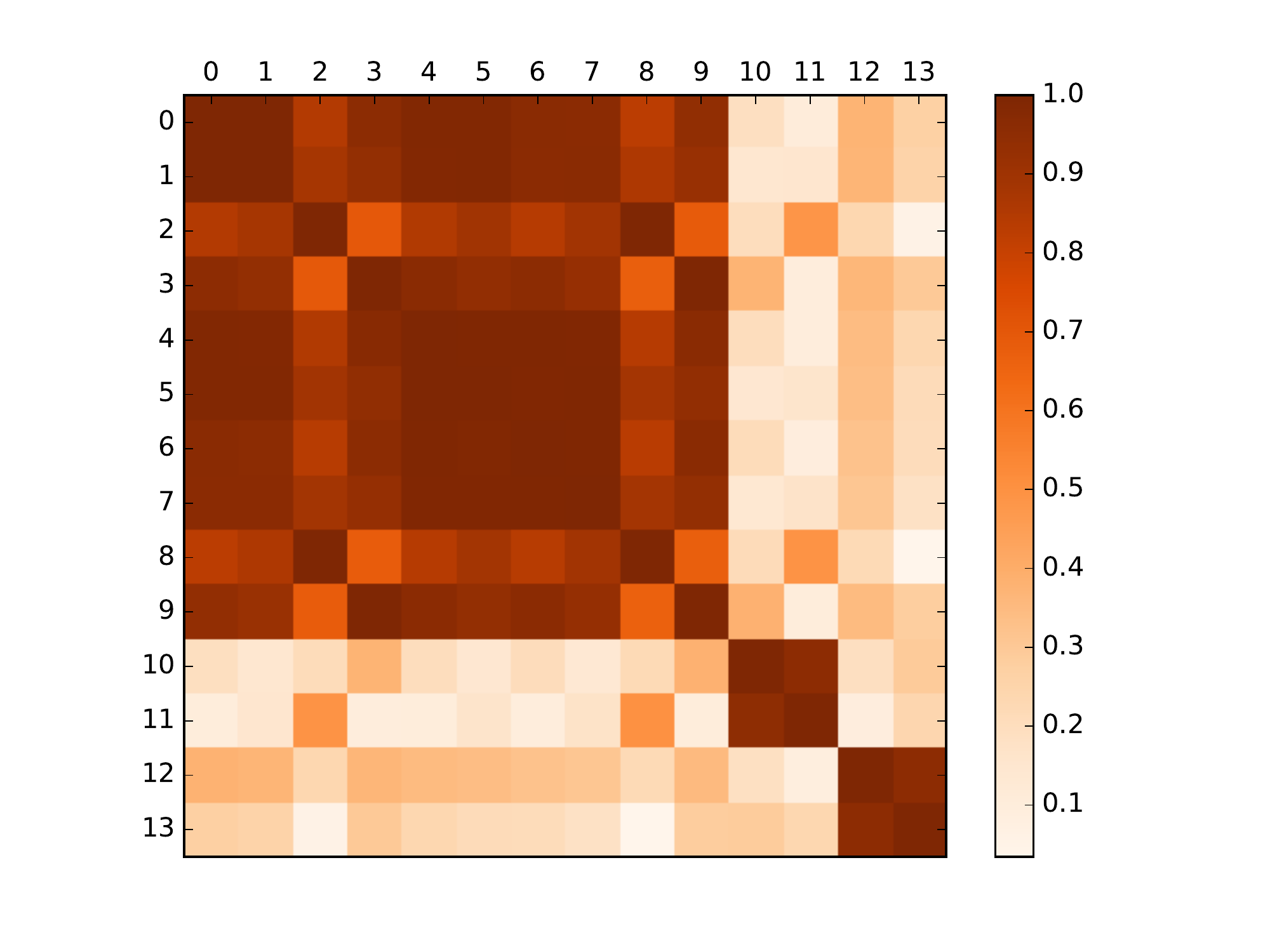}}
\subfigure[Scalar channel]{\includegraphics[width=0.55\textwidth,clip]{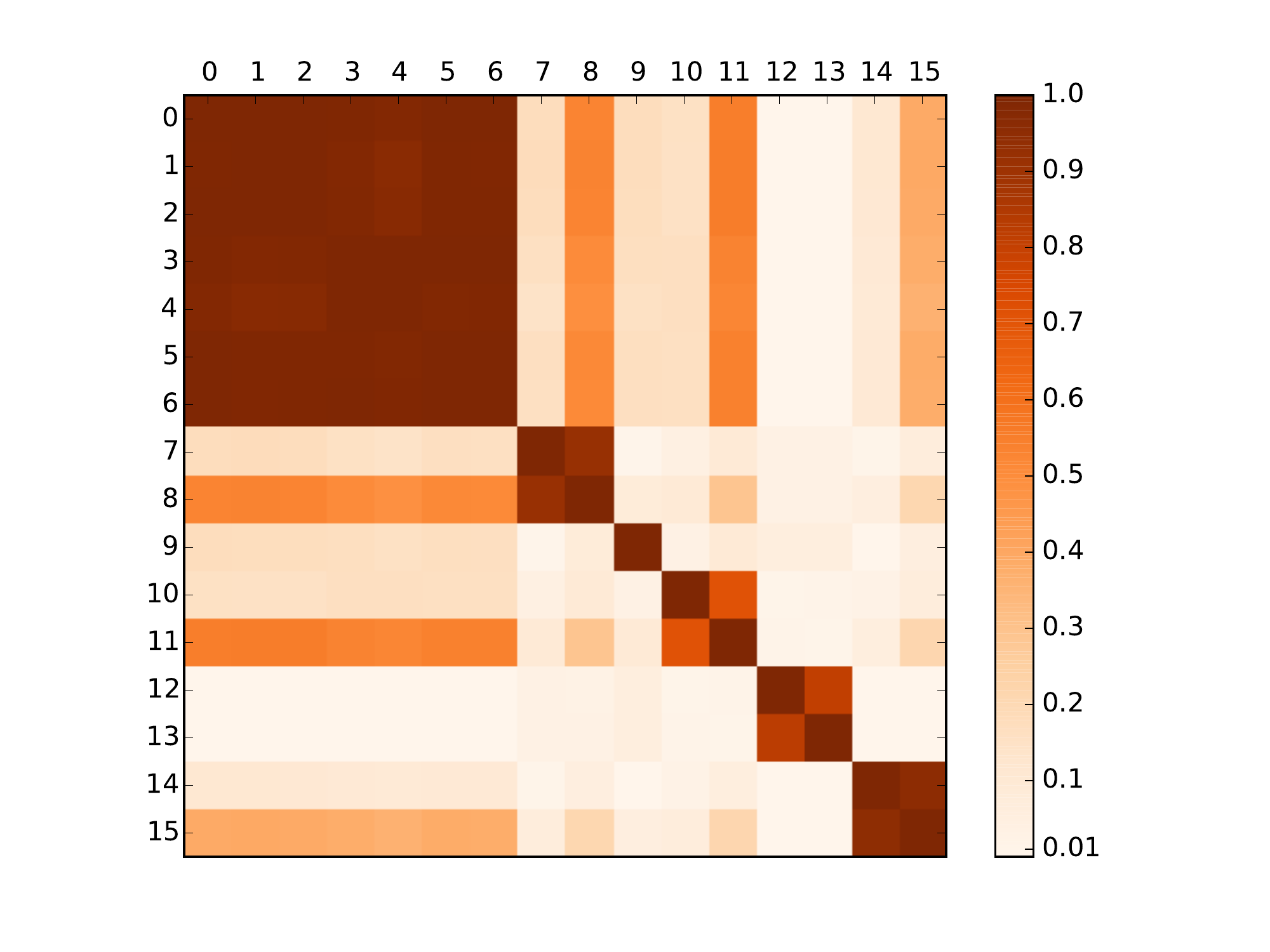}}
\caption{Matrix plots of the normalized correlation matrix $M_{ij}(t)$ in Eq.~(\ref{pearson}) at timeslice $t=5a$ for all operators in the $1^{--}$ (left) and the $0^{++}$ (right) channels from table~\ref{operators}.}\label{fig:c_matrix_plots}
\end{figure}

Once the Laplacian eigenvectors are available, operators and
perambulators can be computed in parallel. So far we have implemented
rest frame operators, see the list in Tab.~\ref{operators}.  Operators
relevant for moving frames are being developed~\cite{sasa_proceeding}
and will be employed in the near future. The energy levels are finally
computed from the solution of the generalized eigenvalue problem of
the correlation matrix
\begin{equation}
 C_{ij}(t) = \langle \mathcal{O}_i(t) \mathcal{O}_j^\dagger (0) \rangle\,,
\end{equation}
as the result of the fit to an exponential decay of the eigenvalues 
\begin{equation}
	\lambda_m(t,t_0) \propto e^{- (t-t_0)E_m} (1+\mathcal{O}(e^{- (t-t_0) \Delta E_m }))  \label{eigv}
\end{equation}
from
\begin{equation}
 C(t) v_m(t,t_0) = \lambda_m(t,t_0) C(t_0) v_m(t,t_0)\,.
\end{equation}
We fix $t_0 = 2$ and we optimize the choice of the basis of operators
by looking at the normalized correlation matrix
\begin{equation}\label{pearson}
 M_{ij}(t) = \frac{C_{ij}(t)}{\sqrt{C_{ii}(t)C_{jj}(t)}}\,.
\end{equation}
The structure of the matrix $M$ can help identify how the operators
are correlated and how different states can be effectively separated
from each others by a suitable choice of the operator basis. From
Fig.~\ref{fig:c_matrix_plots}, in the vector channel we do not see
strong correlations between two- and single-particle operators,
while a signal for stronger correlations appears in the scalar
channel.

Translational invariance in the time direction is broken on
gauge-field configurations with open boundary conditions. Correlators
must be computed with source positions at least 28 timeslices away
from the boundaries, a value chosen by looking for boundary effects in
the pion correlator. In addition, we average on each configuration the
correlators for eight different timesources, well separated in order to
reduce the autocorrelation. Finally, two consecutive analyzed
configurations are separated by twenty molecular dynamics units and
ten accept/reject Metropolis steps.

\section{Results}\label{results}


\begin{figure}[t]
\centering
\subfigure[$\kappa_c = 0.125220$]{\includegraphics[width=.49\textwidth]{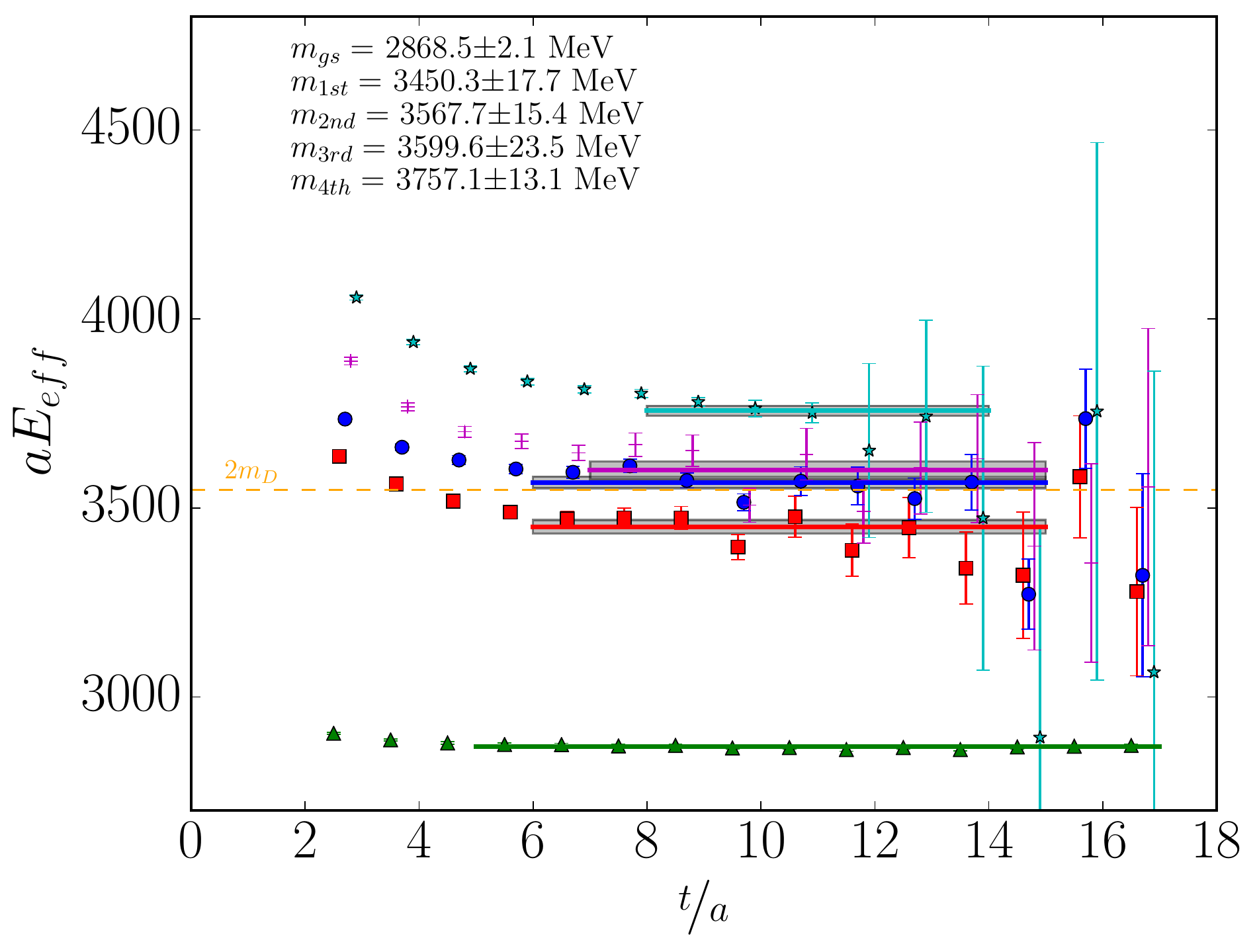}}
\subfigure[$\kappa_c = 0.123147$]{\includegraphics[width=.49\textwidth]{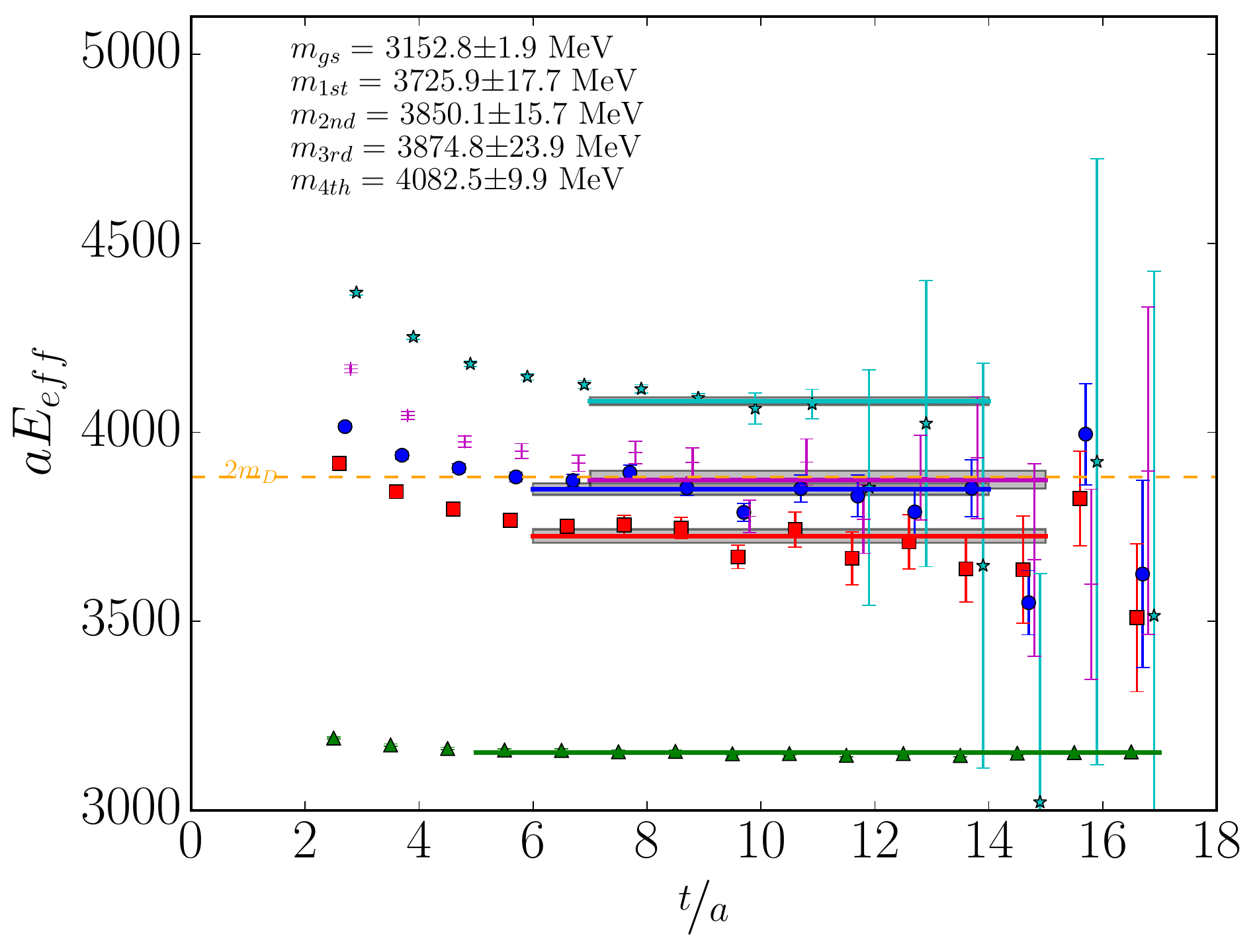}}
\caption{Effective masses of the 5 lowest states in the $1^{--}$ channel with $\kappa_c = 0.125220$ and $\kappa_c = 0.123147$, extracted with the variational analysis method from a basis of 10 operators $\{0,1,4,5,7,8,10,11,12,13\}$. The results are obtained from 125 configurations of the U101 ensemble.}
\label{fig:plot_effmass_T1MMx}
\end{figure}%

The analysis of the two ensembles is not yet completed, therefore at
the moment we can present only the effective energy levels of the
ensemble U101 for total momentum zero.  Our plans for the near future is to double the
statistics on the U101 ensembles and to perform all contractions including
operators in moving frames; (full) distillation is used in this case.  An analogous study is planned on ensemble H105 with 
larger volume, where the perambulators are already computed with stochastic
distillation \cite{Morningstar:2011ka}. The final analysis will also have
to consider the impact of the reweighting factors of the positive twisted mass
introduced to stabilize the HMC trajectory, which have not yet been included. Finally, the finite volume energies will be
computed, and the scattering amplitudes will be suitably parameterized. The main challenge will be the understanding of how the
various decay channels are coupled together for the $0^{++}$
resonances.

Nevertheless, we can already discuss the structure of the discrete
spectrum of energy levels and the impact of tuning the charm quark
mass. In the $1^{--}$ channel, we are able to reliably extract five
energy levels in total, see Fig.~\ref{fig:plot_effmass_T1MMx}. The
entire charmonium spectrum shifts as expected if the charm quark mass
is decreased. The comparison of the two different $\kappa_c$ reveals 
that $E_3-2m_D$  is 
bigger on $k_c=0.125220$ than on $k_c=0.123147$, where $E_3$ is related to $\psi(3770)$. 
It is therefore more likely that $\psi(3770)$ is a resonance above threshold in the former case.  
 The statistical significance is still not high enough to draw a complete
conclusion on this point, and the determination of the error is still
preliminary. It is however clear that the systematic uncertainty of
the tuning of the $\kappa_c$ does not dramatically affect the physical
picture that we aim to investigate, but that a clever choice of the
charm quark mass can help to move the $\psi(3770)$ toward a resonance
state.

\begin{figure}
\centering
\subfigure[$\kappa_c = 0.125220$ excluding strange]{\includegraphics[width=.49\textwidth]{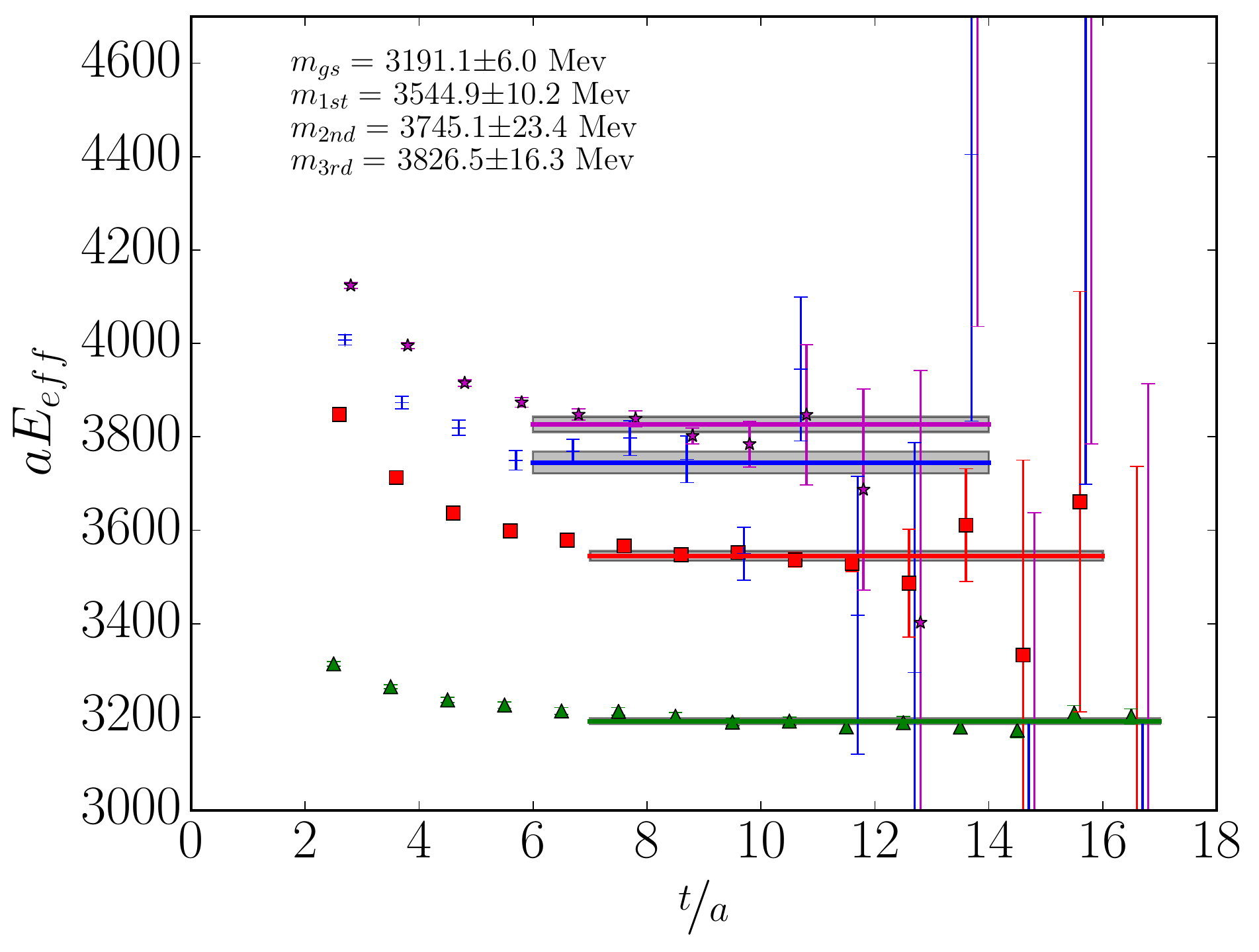}}
\subfigure[$\kappa_c = 0.125220$ including strange]{\includegraphics[width=.49\textwidth]{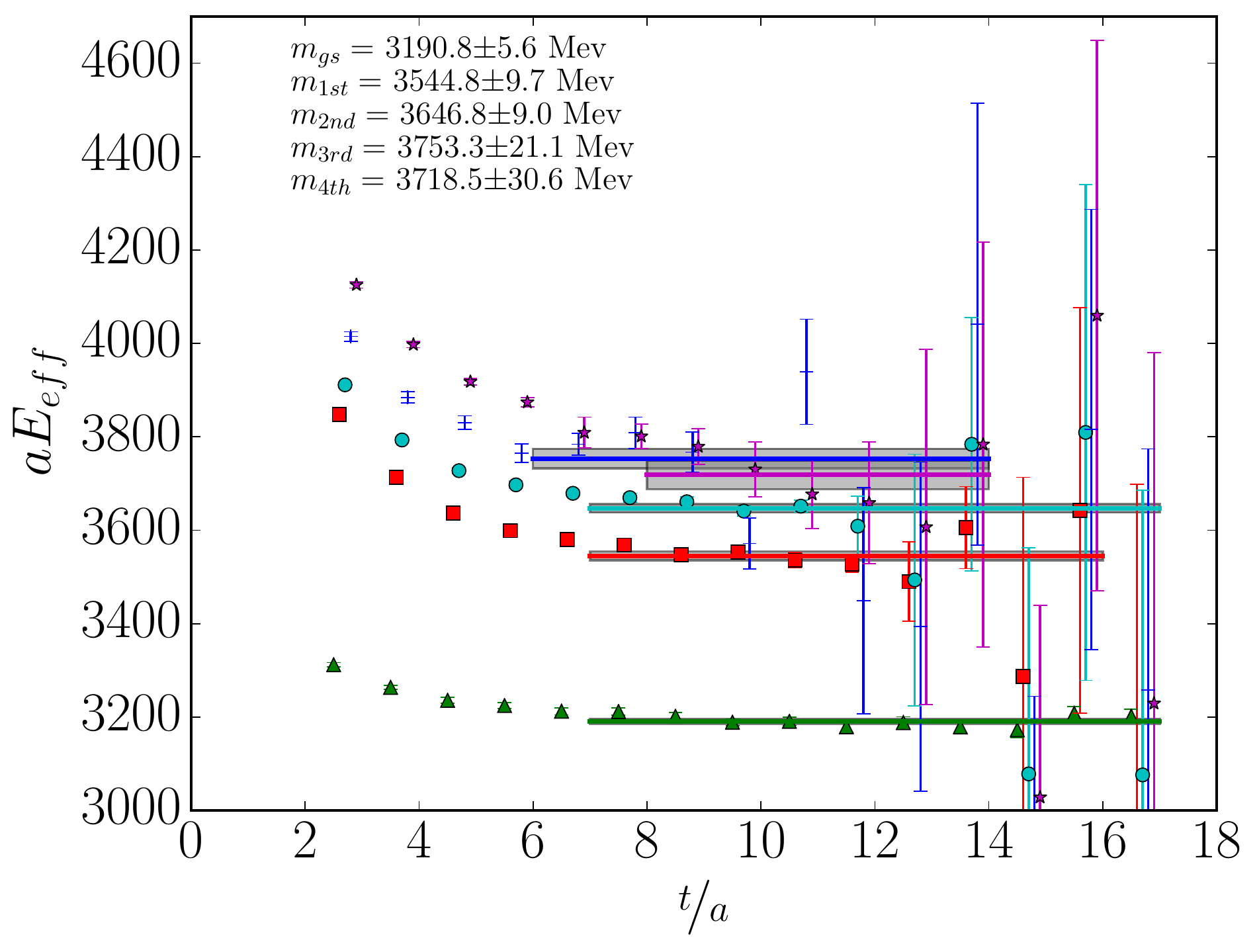}}
\caption{Effective masses of the 4 lowest states in the $0^{++}$ channel with $\kappa_c = 0.125220$, extracted with the variational analysis method from an operator basis excluding (left) and including (right) scattering in mesons with strange valence quarks. The results are obtained from 125 configurations of the U101 ensemble.}
\label{fig:plot_effmass_A1PP}
\end{figure}

The energy levels in the $0^{++}$ channel are presented in
Fig.~\ref{fig:plot_effmass_A1PP}. As expected, the inclusion of hidden
strange sector operators in our basis results in an additional
scattering energy level above the one given by $\bar{D}D$ in s-wave at
zero momentum. At the physical point, the two thresholds of $\bar{D}_sD_s$
and $\bar{D}D$ meson scattering are well separated by approximatively
200 MeV. In our setup, extrapolating along the constant $2m_l + m_s$
line, the gap between the two corresponding scattering levels is
reduced to about 100 MeV. This effect is driven by the fact that we
are working with a heavier than physical light quark mass and an
unphysically light strange quark mass, therefore the resulting
analysis of the scattering amplitude will necessarily have to take into account
the additional scattering channels given by strange mesons.

%
%

\section{Conclusions}
\label{conc}

We reported on the non-perturbative lattice determination of the
properties of the  resonances in the charmonium spectrum. We
studied the impact of the tuning of the charm quark mass on the
lattice energy levels, finding evidences that modifying  $\kappa_c$
might be crucial to move the position of the $\bar{D}D$ threshold
below the second exited states in the vector channel. The preliminary
results of the energy spectrum show that our target statistics are sufficient
to reliably estimate the scattering states relevant for extracting the
scattering amplitudes. The calculations of the contractions on the remaining
configurations as well as the analysis itself is ongoing. The plan for the near future is to provide an analysis
of scalar and vector charmonium resonances on the two ensembles U101
and H105, including scattering in moving frames.

\section{Acknowledgements}
Our code that implements the distillation method is written within the
framework of the Chroma software package~\cite{Edwards:2004}. The
inversion of the Dirac operator is performed used the multigrid solver
of Ref.~\cite{Heybrock:2014iga,Heybrock:2015kpy,Richtmann:2016kcq,Georg:2017diz}. The
simulations were performed on the Regensburg iDataCool cluster, and the SFB/TRR 55 QPACE~2~\cite{Arts:2015jia} and QPACE~3 machines. We thank our colleagues in CLS for the joint effort in the generation of the gauge field ensembles 
which form a basis for the here described computation. The Regensburg group was supported by the Deutsche Forschungs-
gemeinschaft Grant No. SFB/TRR 55. M. P. acknowledges support from EU under grant no. MSCA-IF-EF-ST-744659 (XQCDBaryons).

\end{document}